\documentclass[twocolumn,showpacs,preprintnumbers,amsmath,amssymb]{revtex4}
\usepackage{graphicx}% Include figure files
\usepackage{dcolumn}% Align table columns on decimal point
\usepackage{bm}% bold math
\newcommand{\bea}{\begin{eqnarray}}
\newcommand{\eea}{\end{eqnarray}}
\newcommand{\be}{\begin{equation}}
\newcommand{\ee}{\end{equation}}
\def\shiftleft#1{#1\llap{#1\hskip 0.04em}}
\def\shiftdown#1{#1\llap{\lower.04ex\hbox{#1}}}
\def\thick#1{\shiftdown{\shiftleft{#1}}}
\def\b#1{\thick{\hbox{$#1$}}}
\begin{document}

\title{ {\Large
Charge form factors and nucleon shape}}
\thanks{published in: AIP Conf. Proc. {\bf 904} (2007) pg.~110.} 
%Shapes of Hadrons, Athens, Greece, 27-29 April 2006,
%eds. C.N. Papanicolas and A.M. Bernstein

\pacs{11.30j, 13.40.Gp, 13.40.Em, 13.60.Rj, 14.20.Gk}
%\keywords {electroexcitation of nucleon resonances, spin-flavor symmetry, 
%nucleon intrinsic quadrupole form factor}

\author{A. J. Buchmann}
\email{alfons.buchmann@uni-tuebingen.de} 
\affiliation{Institute for Theoretical Physics \\
University of T\"ubingen \\ 
Auf der Morgenstelle 14 \\  
D-72076 T\"ubingen, Germany}

\begin{abstract}
To obtain further information on the geometric shape of the nucleon,
the proton charge form factor is decomposed into two terms, which are 
connected respectively with a spherically symmetric and an intrinsic 
quadrupole part of the proton's charge density. 
Quark model relations are employed to 
derive expressions for both terms. In particular,  
the proton's intrinsic quadrupole form factor 
is obtained from a relation between the $N\to \Delta$ 
and neutron charge form factors. 
The proposed decomposition shows that the neutron charge form factor 
is an observable manifestation of an intrinsic quadrupole form factor 
of the nucleon. Furthermore, it affords an interpretation of recent 
electron-nucleon 
scattering data in terms of a nonspherical distribution of quark-antiquark 
pairs in the nucleon.
\end{abstract}

\maketitle

%%%%%%%%%%%%%%%%%%%%%%%%%%%%%%%%%%%%%%%%%%%%
%% MAINMATTER
%%%%%%%%%%%%%%%%%%%%%%%%%%%%%%%%%%%%%%%%%%%%

\section{Electromagnetic probing of nucleon structure} 

\begin{figure*}[htb]
\includegraphics[height=0.209\textheight]{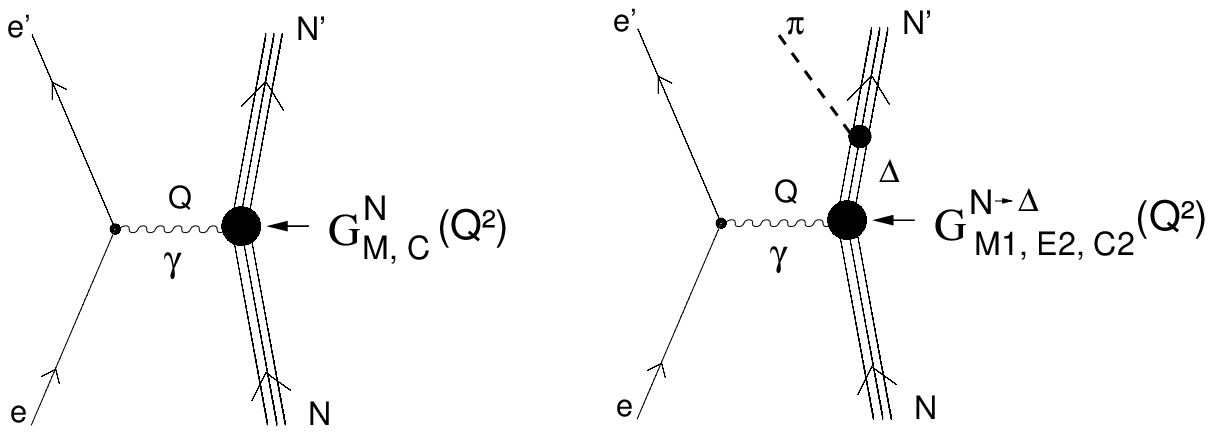}
\caption{\label{figure:scattering} 
Left: Probing of nucleon structure via elastic electron-nucleon 
scattering $e\, N \to e' \, N'$ 
involving the exchange of a single virtual photon $\gamma$ of 
momentum $Q$. 
The nucleon structure information is contained in 
the charge monopole form factor $G^{N}_{C}(Q^2)$ and
magnetic dipole form factor
$G^{N}_{M}(Q^2)$.
Right: Inelastic electron-nucleon scattering $e \, N \to e' \, \Delta$. 
The excitation of the $\Delta$ resonance is described 
by three electromagnetic transition form factors 
$G^{N \to \Delta}_{M1}(Q^2)$, $G^{N \to \Delta}_{E2}(Q^2)$, 
and $G^{N \to \Delta}_{C2}(Q^2)$. 
Relations between elastic and inelastic nucleon charge form factors 
provide information on the spatial shape of the nucleon.}
\end{figure*}

Elastic electron-proton scattering 
experiments done at the Stanford Linear Accelerator about 
50 years ago have shown that the proton has a finite size of about
0.9 fm. In addition, these and subsequent 
experiments have provided detailed information 
on the radial variation of the charge and magnetization densities 
of the proton~\cite{cha56,sim80,kel02,fri03}. 
Nucleon structure information is encoded in two elastic electromagnetic 
form factors, which parametrize the deviation of the measured 
cross section from the theoretical Mott cross section $\sigma_M$. The latter
describes the scattering of a pointlike spin 1/2 electron on a spinless and 
structureless nucleon (Fig.~\ref{figure:scattering}). 
In the one-photon exchange approximation, the differential cross section  
for elastic electron-nucleon scattering, where only electrons scattered 
into solid angle $d\Omega$ are detected, and where projectile and target are 
unpolarized  can be written as~\cite{hal84}   
\bea
\label{elasticcrosssection}
\left (\frac{d\sigma}{d\Omega}\right )_{el} & = &
\sigma_{M} \, f_{rec} \, \Biggl \{ \frac{1}{1+\tau} \left ( {G_{C}^N}^2(Q^2) + 
\tau \, {G_{M}^N}^2(Q^2)  \right ) \nonumber \\
& &  \ \ \ \ \ \ \ \ \ \ \, \, \ + \, 2\, \tau\, 
\tan^2 \left (\frac{\Theta}{2}\right ) 
{G_{M}^N}^2(Q^2)  \Biggr \}.
\eea
This equation is known as Rosenbluth formula. Here,  $G_{C}^N(Q^2)$ 
and $G_{M}^N(Q^2)$ are the charge monopole and magnetic dipole
form factors of the nucleon, usually refered to as Sachs 
form factors~\cite{footnote1}.
The square of the four momentum transfer of the virtual photon~\cite{comment1} 
is denoted by $Q^2$ and $\tau=:Q^2/(4\, M_N^2)$, where $M_N$ is the nucleon
mass.  Furthermore, 
$f_{rec}= ( 1 + 2(\epsilon_i/M_N) \, \sin^2(\Theta/2))^{-1}$ is the nucleon 
recoil factor, where $\epsilon_i$ is the incident electron energy, 
and $\Theta$ is the electron scattering angle. 

The form factors in Eq.(\ref{elasticcrosssection}) can be determined from the
measured cross section with the help of the Rosenbluth 
separation method~\cite{comment2}.
Unlike the cross section, which contains
coordinate frame dependent kinematical variables, the form factors depend
only on the Lorentz invariant $Q^2$. 
They are therefore well suited for a comparison between experiment and theory.
At $Q^2=0$ the Sachs form factors 
are normalized to the charge 
and magnetic dipole moment of the nucleon, i.e., 
$G_{C}^p(0)=1$ and $G_{M}^p(0)=\mu_p$ for the proton, and 
$G_{C}^n(0)=0$ and $G_{M}^p(0)=\mu_n$ for the neutron.

Until a few years ago, unpolarized scattering data analyzed with 
the Rosenbluth separation method provided ample
evidence for a scaling law, i.e.,  for a wide range
of momentum transfers both proton form factors, and the neutron
magnetic form factor could  be approximately
described by a common dipole function $G_D(Q^2)$ 
\be
\label{scaling}
G^p_{C}(Q^2) = \frac{G^p_{M}}{\mu_p}(Q^2)  = \frac{G^n_{M}}{\mu_n}(Q^2)  
= G_D(Q^2)
%= \left (\frac{1}{1+Q^2/ \Lambda^2} \right )^2, 
\ee
where  $G_D(Q^2) = ({1+Q^2/ \Lambda^2})^{-2}$, and 
$\Lambda^2$ is related to the spatial extension of the proton charge
and magnetization densities. 
These equations suggest that at each point in space
the local charge and magnetization densities
in the proton are approximately equal, and that
the magnetization density in the proton is equal to the 
magnetization density in the neutron. 
The neutron charge form factor has a different functional behavior,
and will be discussed below.
It has been shown that a scaling law 
as in Eq.(\ref{scaling}) arises in constituent quark models if exchange 
currents between the quarks are neglected, and if the charge and magnetic
form factors of constituent quarks are assumed to be equal~\cite{mor99}.

%%%%%%%%%%%%%%%%%%%%%%%%%%%%%%%%%%%%%%%%%%%%%%%%%%%%%%%%%%%%%%%%%%%%%%%%%
%             FIG. 1:Elastic and Inelastic electron nucleon scattering
%%%%%%%%%%%%%%%%%%%%%%%%%%%%%%%%%%%%%%%%%%%%%%%%%%%%%%%%%%%%%%%%%%%%%%%%%
%%%%%%%%%%%%%%%%%%%%%%%%%%%%%%%%%%%%%%%%%%%%%%%%%%%%%%%%%%%%%%%%%%%%%

More recent electron-proton experiments employing polarized 
electrons and measuring the polarization of the recoiling proton 
in coincidence with the scattered 
electron have led to very different results for the nucleon form 
factors~\cite{jon00,comment7,Gui03}. Unlike the unpolarized elastic cross 
section of Eq.(\ref{elasticcrosssection}), which is proportional to an 
incoherent sum of the squares of these form 
factors, the polarized cross section also contains interference terms of 
the two elastic form factors. The latter permit the extraction of more 
accurate results for the charge form factor compared to those obtained 
with Rosenbluth separation. For a review see Ref.~\cite{arr03}.
At high momentum transfers the new experiments on the proton 
show large deviations from scaling between charge and magnetic form factors. 
This is indicative of substantial differences 
between the charge and magnetization distributions of the proton.
Thus, the elastic form factors contain more information
on nucleon structure than previously thought. 
In the following we suggest that the scaling law violation 
is a consequence of the proton's nonspherical charge distribution.

First evidence for a deviation of the nucleon's
charge distribution from spherical symmetry came from studies of 
inelastic electron-nucleon scattering (Fig.~\ref{figure:scattering}) 
in which the $\Delta(1232)$ 
resonance with spin 3/2 and isospin 3/2 is excited. The $\Delta$ resonance 
is the lowest lying excited state of the nucleon $N(939)$ with the same 
quark content as the ground state.  
Parity invariance and angular momentum conservation
restrict the electromagnetic $N \to \Delta$ excitation
to magnetic dipole (M1), electric quadrupole (E2), and charge
(or Coulomb) quadrupole (C2) transitions. In terms of 
transition multipole form factors, which are the inelastic counterparts of the
elastic Sachs form factors in Eq.(\ref{elasticcrosssection}), 
the cross section for electroproduction of the $\Delta$-resonance is 
given by a generalized Rosenbluth formula
\bea
\label{inelasticND}
\left (\frac{d\sigma}{d\Omega}\right )_{inel} 
\!\!\!\!\!\!\!\!\!\!\!\! & = &   
\!\!\! \sigma_{M} \, f_{rec} \,\frac{Q^2}{4M_N M_{\Delta}}\, 
%\frac{Q^2}{{\bf q}^2} 
\frac{1}{1+\tilde{\tau}} 
\Biggl \{ \frac{Q^2 \,M_N^2}{9}\, ({G_{C2}^{N\to \Delta}})^2(Q^2) \nonumber \\
& & \hspace{-1.0cm}  + 
\epsilon^{-1} \left ( ({G_{M1}^{N \to \Delta}})^2(Q^2)  + 
\frac{M_N^2 {\bf q}^2}{18}\,({G_{E2}^{N\to \Delta}})^2(Q^2) \right )
\Biggr \}, 
\eea
where $\epsilon = (1+ 2\, (M_{\Delta}^2/{M_N^2})  
(\vert {\bf q} \vert^2/Q^2)  \tan^2 (\Theta/2))^{-1}$ is the polarization of 
the transverse photons in the laboratory frame, and 
$\tilde{\tau}=:Q^2/(M_N+M_{\Delta})^2$. The form factors are normalized to the 
transition magnetic moment $G_{M1}^{N \to \Delta}(0)=\mu_{N \to \Delta}$ 
and quadrupole moment $G_{C2}^{N \to \Delta}(0)=Q_{N \to \Delta}$. 
Cross section formulae analogous to Eq.(\ref{inelasticND}) have been 
written by several authors~\cite{jon73,car86}. 
The generalized Sachs form factors in Eq.(\ref{inelasticND}) are based 
on the same definition of the multipole operators~\cite{def66} used 
for the elastic form factors in Eq.(\ref{elasticcrosssection}), 
and facilitate the comparison between elastic and inelastic nucleon 
form factors. 

In the cross section of Eq.(\ref{inelasticND}) the pions from the strong decay
of the $\Delta$-resonance are not observed,  
only the scattered electrons are detected.  As a result, the 
cross section formula involves only an incoherent sum of the squares of the 
multipole form factors, and the small quadrupole form factors cannot be 
reliably determined. On the other hand, if the nucleon or 
pion from the decay $\Delta \to N + \pi$ 
is detected in coincidence with the scattered electron, 
as depicted in Fig.~\ref{figure:scattering}, the corresponding
cross section formula contains an interference term between the small 
charge quadrupole and the large magnetic dipole amplitudes. The contribution
of the former to the cross section is thereby greatly enhanced, and can be more
reliably determined. For a discussion of the coincidence 
cross section for pion electroproduction see Refs.~\cite{Ber03}. 

At low momentum transfers, experiment shows that the $N \to \Delta$ excitation 
is predominantly an M1 transition, which can be interpreted as the spin 
and isospin flip of a single quark. The quadrupole amplitudes are only about 
$1/40$ of the dominant magnetic dipole amplitude. They can be interpreted as 
arising from a double spin flip involving two interacting quarks~\cite{Buc97}. 
Despite their smallness, the C2 and E2 multipoles have been the focus of many 
recent experimental~\cite{Ber03,Buu02,Joo02,Bar02,sta06,els06,spa05,ung06} 
and theoretical works~\cite{Idi04,Tia03,Jen02,Ale03,Hes02,gro06,pas06}.
It becomes clear below that the transition quadrupole moments are nonzero 
only if the geometric shape of the nucleon deviates from spherical symmetry, 
and that from their sign and size information on the shape of the 
nucleon's charge distribution can be infered~\cite{Hen01}. 

In this paper we show that further information on the 
geometrical shape of the nucleon can be obtained from the study of 
the relations between inelastic and elastic form factors. Generally, 
the geometric properties of 
the ground state of a physical system, in particular its size and shape,
have a direct bearing on the eigenfrequencies and eigenmodes of its 
excitation spectrum. Conversely, knowledge of the eigenfrequencies
and excitation modes of a system enables us to draw certain conclusions 
concerning its size and shape. In the case of the nucleon, we have suggested 
that the existence of a quadrupole excitation mode of the $\Delta$-resonance 
is closely related to a quadrupole deformation of the nucleon's  
charge distribution as reflected by a positive intrinsic quadrupole 
moment~\cite{Hen01} and an intrinsic 
charge quadrupole form factor~\cite{buc05}. 
The latter has observable consequences for the elastic nucleon form factors. 
To show this,  we decompose in chapter~\ref{cha:4} 
the proton charge form factor into a spherically 
symmetric term and a nonspherical intrinsic quadrupole term, where the 
latter is given by the $N \to \Delta$ charge quadrupole transition 
form factor. Before doing this, we review in chapter~\ref{cha:2} 
the empirical information on the relations between the elastic nucleon and 
inelastic $N \to \Delta$ electromagnetic form factors,
and how these can be understood in terms of strong interaction symmetries, 
in particular spin-flavor symmetry. 

\section{Electromagnetic $N \to \Delta$ transition and spin-flavor symmetry}
\label{cha:2}
\subsection{Strong interaction symmetries}

It is well known that invariance of the strong interaction under SU(2) isospin 
transformations leads to isospin conservation and the appearance 
of degenerate hadron multiplets with fixed isospin, in which individual 
hadrons  have different charges but nearly the same mass. 
Moreover, also other properties of the particles within an isospin multiplet 
are closely related by the underlying symmetry 
group. This remains true even if the symmetry is broken. 
As discussed below, the group algebra guarantees that symmetry breaking 
occurs according to a well defined scheme, with the result that its 
consequences for different members of an isospin multiplet are related by 
the Wigner-Eckart theorem, i.e., the matrix elements of a given isospin 
breaking operator evaluated for different members of an isospin multiplet 
differ only by a Clebsch-Gordan coefficient.

But strong interactions are invariant under a higher symmetry than isospin.  
Flavor SU(3) symmetry ties together 
isospin multiplets with different isospin and different strangeness 
to larger multiplets with the same spin and parity, 
e.g., octets and decuplets~\cite{Gel64}. 
Even though mass differences between different isospin multiplets within 
the baryon flavor octet and decuplet are of the order of 100 MeV, the 
inclusion of the relevant symmetry breaking operators leads to a number of 
remarkable predictions, such as the Gell-Mann-Okubo relation between octet 
baryon masses, or the equal spacing rule between the masses of the 
isospin multiplets in the baryon decuplet, which are in 
good agreement with experiment.

An even higher strong interaction symmetry than SU(3) flavor is SU(6) 
spin-flavor, which unites the spin 1/2, flavor octet baryons 
($2 \times 8$ states), among them the familiar proton and neutron, and the 
spin 3/2, flavor decuplet baryons ($4 \times 10$ states), among them the four 
$\Delta$ states into a common {\bf 56}-dimensional 
supermultiplet~\cite{Gur64,Sak64,Beg64}. More formally, 
the SU(6) ground state supermultiplet can be decomposed into irreducible 
representations of the flavor and spin subgroups as
${\bf 56} = ({\bf 8}, {\bf 2}) +  ({\bf 10}, {\bf 4})$, where 
the first and second label refer to the SU(3) and SU(2) dimensions
respectively.
There are numerous successful predictions based on SU(6) spin-flavor 
symmetry. For example, SU(3) flavor symmetry alone does not suffice to
uniquely determine the ratio of proton and neutron magnetic moments, 
whereas SU(6) spin-flavor symmetry leads to the prediction~\cite{Beg64}  
$\mu_p/\mu_n=-3/2$, in excellent agreement with the experimental result
-1.46.  Another example for the predictive power of SU(6) 
spin-flavor symmetry is the G\"ursey-Radicati mass formula~\cite{Gur64},
which explains why the Gell-Mann Okubo mass formula works so well for 
both octet and decuplet baryons with the same numerical coefficients.
Without SU(6) it would remain a  mystery why the violation of 
SU(3) symmetry in the baryon octet is the same as in the baryon decuplet. 
This fact is not explained in SU(3). 

Thus, SU(6) is an excellent symmetry in baryon 
physics, and the question arises whether it 
is a symmetry of quantum chromodynamics. This is indeed the case. 
In an $1/N_c$ expansion of QCD,  where $N_c$ denotes the number of colors,  
it has been shown that QCD possesses a 
spin-flavor symmetry, which is exact
 in the large $N_c$ limit~\cite{sak84,das95},
and that for finite $N_c$ spin-flavor symmetry 
breaking operators can be classified 
according to the powers of $1/N_c$ associated with them.
It turns out that higher orders of spin-flavor symmetry breaking 
are suppressed by correspondingly higher powers of $1/N_c$. 
For example, second and third order SU(6) symmetry breaking
described by two- and three-quark operators
are suppressed by $1/N_c$ and $1/N_c^2$ respectively, compared 
to the first order symmetry breaking due to one-quark operators. 
As a result, one obtains a rigorous perturbative expansion scheme 
for QCD processes that works at all energy scales, and which furthermore 
provides a connection between broken SU(6) spin-flavor symmetry 
and the underlying quark-gluon dynamics~\cite{leb98}. 
This allows us to put certain model results on a solid theoretical foundation.
In the next section, we employ SU(6) symmetry and its breaking
to study the relations between inelastic $N \to \Delta$ transition form 
factors and elastic nucleon form factors.

\subsection{Electromagnetic $N \to \Delta$ transition and nucleon form factors}

Because the $N$ and $\Delta$ belong to the same ${\bf 56}$-dimensional
ground state multiplet of the SU(6) spin-flavor group their
properties are related. In particular, the electromagnetic $N \to \Delta$
transition form factors are related to the electromagnetic elastic form
factors of the nucleon.  In fact, the SU(6) relation between the 
magnetic dipole transition form factor 
$G^{N\to \Delta}_{M1}(Q^2)$ and the elastic neutron magnetic
form factor $G_M^n(Q^2)$ has been known for some
time~\cite{Beg64}
\bea
\label{ffrel1}
G_{M1}^{N \to \Delta}(Q^2) &  = &  - \sqrt{2} \, \, G_M^n(Q^2).
\eea
At $Q^2=0$, both form factors are normalized to their
magnetic dipole moments $\mu_{N \to \Delta}$ and $\mu_n$
\be
\label{ffrel1stat}
\mu_{N \to \Delta} = - \sqrt{2} \, \, \mu_n.
\ee
These relations also hold when second order SU(6) symmetry
breaking operators are included~\cite{Leb95}. They have also been derived
in the quark model with two-quark currents~\cite{Buc00}, and
are violated only by three-quark currents~\cite{Dil00} or
third order SU(6) symmetry breaking operators~\cite{Leb04}.
The latter are suppressed by a
factor $1/N_c^2$ with respect to the leading term so that these
relations are valid in good approximation.

The other relation between the charge
quadrupole transition form factor $G^{N\to \Delta}_{C2}(Q^2)$ and the
elastic neutron charge form factor $G_C^n(Q^2)$
\bea
\label{ffrel2}
G_{C2}^{N \to \Delta}(Q^2) &  = &  -\frac{3\,\sqrt{2}}{Q^2} G_C^n(Q^2)
\eea
has been found only quite recently within a quark model that includes 
in addition to the usual single quark electromagnetic current also two-body 
exchange currents associated with the quark-quark interaction~\cite{Buc00}. 
In the $Q \to 0$ limit, Eq.(\ref{ffrel2}) reduces to
a relation~\cite{Buc97} between the transition quadrupole
moment $Q_{N \to \Delta}$ and the neutron charge radius
$r_n^2$
\be
\label{ffrel2stat}
Q_{N \to \Delta}= \frac{1}{\sqrt{2}} \, r_n^2,
\ee
which is in good agreement with recent extractions of
$Q_{N \to \Delta}$ from the data~\cite{Tia03,Bla01}.
Also at low momentum transfers,
Eq.(\ref{ffrel2}) is satisfied by the data~\cite{Gra01}.  
More recently, it has been shown
that the validity of Eq.(\ref{ffrel2})
is not confined to low momentum transfers but extends into the
 GeV region~\cite{Buc04}.
 
Experimental results are often given for the  $C2/M1$ ratio, which
is defined in terms of the
$N \to \Delta$ transition form factors times a kinematical
factor~\cite{jon73,comment0}
\be
\label{c2m1def}
    \frac{C2}{M1}(Q^2)
= : \frac{\vert {\bf q} \vert \, M_N}{6} \, \,
\frac{G_{C2}^{N \to \Delta}(Q^2)}{G_{M1}^{N \to \Delta}(Q^2)},
\ee
where $M_N$ is the nucleon mass and $\vert {\bf q} \vert $ is
the three-momentum transfer of the virtual photon in the
$\gamma N$ center of mass frame~\cite{comment5}.
Inserting the above form factor relations of Eq.(\ref{ffrel1})
and Eq.(\ref{ffrel2}), the $C2/M1$ ratio can be expressed as the
product of $G_C^n/G_M^n$ and a factor
\be
\label{c2m1ratio}
    \frac{C2}{M1}(Q^2) =
\frac{\vert {\bf q} \vert}{Q} \,  \frac{M_N}{2 Q} \, \,
\frac{G_{C}^n(Q^2)}{G_M^n(Q^2)} =: {\cal R}_n(Q^2).
\ee
We abbreviate this product as ${\cal R}_n(Q^2)$.
Thus, the inelastic $N \to \Delta$ and the elastic neutron form factor ratios
are related.  The theoretical uncertainty of this relation
is mainly due to third order SU(6) symmetry breaking terms
(three-quark currents) omitted in Eq.(\ref{ffrel1}). We estimate it
to be of order $1/N_c^2$ or 10$\%$ (slightly increasing the predicted
$C2/M1$ ratio).
%%%%%%%%%%%%%%%%%%%%%%%%%%%%%%%%%%%%%%%%%%%%%%%
%             FIG. 2: ratio C2/M1
%%%%%%%%%%%%%%%%%%%%%%%%%%%%%%%%%%%%%%%%%%%%%%%
\begin{center}
\begin{figure*}[htb]
\includegraphics[height=.54\textheight]{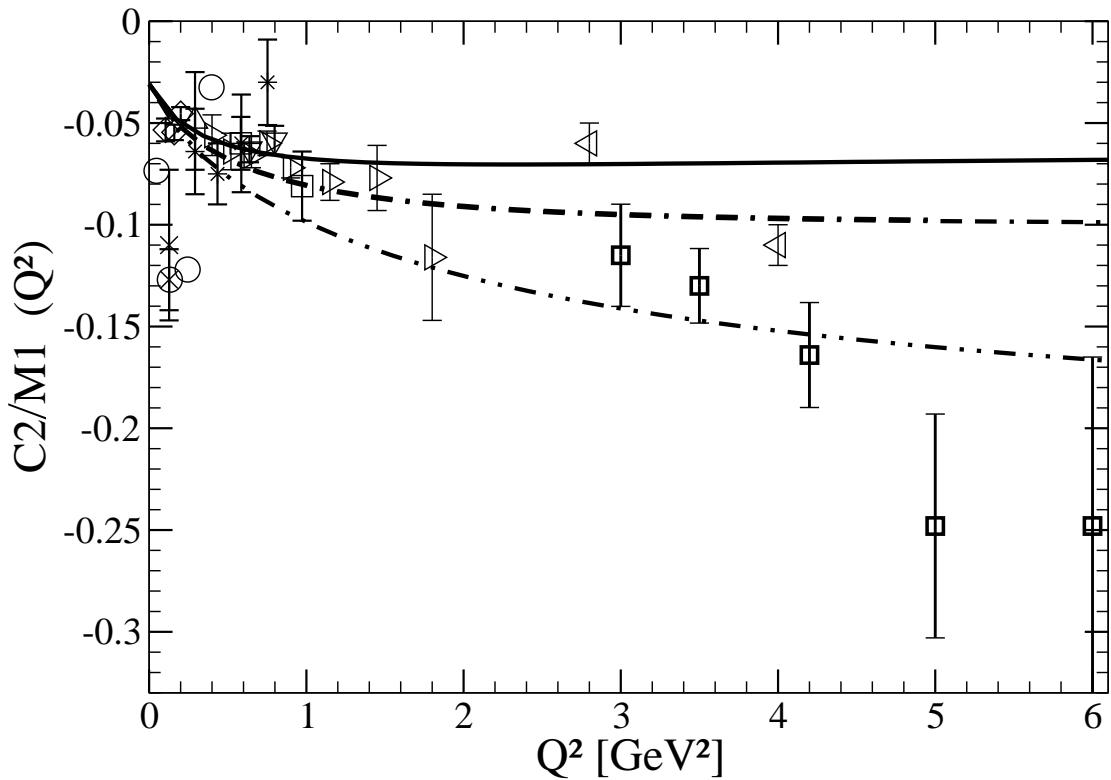}
\vspace{-0.6 cm}
\caption{\label{fig:fig2}
The ratio ${\cal R}_n$ of Eq.(\ref{c2m1ratio})
calculated from a two-parameter fit of
elastic neutron form factor data according to Eq.(\ref{Galster}).
Solid curve for $a=0.9$ and $d=2.8$, dashed-dotted curve for
$a=0.9$ and $d=1.75$~\cite{Gra01}, and double-dotted curve
for $a=0.9$ and $d=0.8$. This is compared with
experimental results for the $C2/M1$ ratio extracted from
pion electroproduction cross 
sections~\protect\cite{Joo02,ung06,Got01,Sid71,Fro98,Ald72,kel05}. }
\end{figure*}
\end{center}
%%%%%%%%%%%%%%%%%%%%%%%%%%%%%%%%%%%%%%%%%%%%%%%%%%%%%%%%%%%%%%%%%%%%%
For convenience, we calculate the ratio ${\cal R}_n$ of Eq.(\ref{c2m1ratio}) 
using for the numerator a two-parameter fit~\cite{Gal71} of the $G_C^n$ data
and for the denominator the dipole fit $G_D$ for $G_M^n$, i.e.,
\bea
\label{Galster}
G_C^n(Q^2) & = &  -\mu_n \, \frac{ a \tau}{1 + d \tau} \, G_D(Q^2), 
\nonumber \\ 
G_M^n(Q^2) &  = & \mu_n \, G_D(Q^2),
\eea
where
$G_D$ is the dipole form factor of Eq.(\ref{scaling}) 
with $\Lambda^2=0.71$ GeV$^2$. 
The $C2/M1$ ratio is then given in terms of the parameters $a$ and $d$,
which have been determined from the lowest moments 
($r_n^2$ and $r_n^4$) of the experimental
neutron charge form factor~\cite{Gra01}. 
In Fig.~\ref{fig:fig2} we plot ${\cal R}_n$
calculated from $G_C^n/G_M^n$ data using Eq.(\ref{Galster})
and compare with $C2/M1$ data from various pion electroproduction
experiments~\cite{Joo02,ung06,Got01,Sid71,Fro98,Ald72}. 
The solid, dashed-dotted, and double dashed-dotted lines correspond
to decreasing values of the Galster parameter $d$.
The chosen parameterization is most likely too restrictive at high $Q^2$,
where it would be preferable to compare directly with measured neutron 
form factor data.

Next, we evaluate Eq.(\ref{c2m1ratio}) for
very low and very high $Q^2$. In the real photon limit 
$Q \to 0$ we obtain~\cite{comment5}
\be
\label{static}
\frac{C2}{M1}(0) \! =  \!
-\frac{M_{\Delta}^2 -M_N^2}{2 M_{\Delta}}\,
\frac{M_N}{12}\, \frac{r^2_n}{ \mu_n} = -0.031
\ee
in good agreement with the experimental $E2/M1$
ratio obtained from pion photoproduction
by different groups~\cite{Bla01,Bec97,comment6,Buc98}.
This result explains the experimental value for the $C2/M1$ ratio
in terms of the charge radius and the magnetic moment of the neutron.
We understand therefore why $C2/M1(0)= -0.03$.
For $Q^2 \to \infty$ we obtain using Eq.(\ref{Galster})
\be
\label{asymp}
{\cal R}_n(Q^2\to \infty)
=\frac{1}{4}\, \frac{M_N}{M_{\Delta}} \left (-\frac{a}{d}\right ),
%=\frac{1}{4}\, \frac{M_N}{M_{\Delta}} \left (-\frac{r_n^2}{2 \mu_n}, 
%\frac{2}{r_{N \to \Delta}^2-12/\Lambda^2}\right )
\ee
which ranges between $-0.06$ and $-0.21$ depending on the parameter $d$.
Thus, we see that the $C2/M1$ ratio asymptotically
approaches a small negative constant determined by the neutron structure
parameters $a$ and $d$. This is in qualitative agreement with expectations
from PQCD~\cite{Idi04} modulo logarithmic corrections. 

We conclude that the two data sets, which were 
thought to be quite independent of each other, satisfy the proposed
relation Eq.(\ref{c2m1ratio}) within experimental uncertainties 
for momentum transfers between 0 and 3 GeV$^2$, which in turn suggests
that Eq.(\ref{ffrel2}) is well satisfied in nature.
As to the physical interpretation of this relation, 
we learn from Eq.(\ref{ffrel2stat}) that the small deviation of $r_n^2$
from zero and the deviation of the nucleon's geometric shape from spherical
symmetry as manifested in a nonzero $Q_{N \to \Delta}$ are closely related
aspects of nucleon structure. Both phenomena have their origin in a
nonspherical cloud of quark-antiquark pairs in the nucleon~\cite{Hen01}, which 
are effectively described as exchange currents between quarks~\cite{Buc91}.  
This interpretation continues to hold for finite momentum transfers as 
suggested by the agreement of Eq.(\ref{c2m1ratio}) with the data. 
We elaborate on this in chapter~\ref{cha:4}.  
On a more abstract level, one can understand this relation 
in terms of the underlying spin-flavor symmetry of QCD and its breaking 
by spin-dependent exchange currents.

\subsection{SU(6) spin-flavor symmetry analysis} 

We start from the observation that the $N(939)$ and $\Delta(1232)$ are members 
of the same {\bf 56} dimensional ground state multiplet of SU(6) spin-flavor 
symmetry. If this symmetry were exact,
$N$ and $\Delta$ baryons would have the same mass, and the form factors
$G_C^n(Q^2)$ and $G_{C2}^{N \to \Delta}(Q^2)$ would be exactly
zero. Spin-dependent two-body potentials in the Hamiltonian break SU(6)
symmetry and lift the degeneracy between $N$ and $\Delta$  
masses~\cite{leb00}.
Similarly, spin-dependent two-body terms in the charge operator
break SU(6) symmetry and lead to nonzero neutron and $N \to \Delta$ 
charge form factors, which are related as in Eq.(\ref{ffrel2}) because 
the group algebra connects the breaking of the symmetry in
$G_C^n$ to the symmetry breaking in $G_{C2}^{N \to \Delta}$. 

This can be most easily seen from a multipole expansion of the relevant 
two-quark charge density $\rho_{[2]}$ in spin-flavor space
\be
\label{su6break}
\rho_{[2]}  =
-B \sum_{i \ne j}^{3} \, e_i \left [ 2 \, \b{\sigma}_i \cdot
\b{\sigma}_j -
(3 \sigma_{i\, z} \sigma_{j\,z}
- \b{\sigma}_i \cdot \b{\sigma}_j) \right ] = 2 \, {\cal S} - {\cal T}, 
\ee
where $B$ represents the color and orbital degrees of freedom,  and
$e_i=(1 + 3\,\tau_{3\, i})/6$ is the quark charge. Here, $\b{\sigma}_i$ and 
$\b{\tau}_i$ are the spin and isospin Pauli matrices of the i-th quark. 
The subscript $[2]$ indicates that this
operator acts on two qarks at a time. As will become clearer below,
to first order flavor (isospin) breaking, 
Eq.(\ref{su6break}) represents the most general two-body charge 
operator in spin-flavor (isospin) space.  The factors multiplying the 
spin scalar ${\cal S}= -B \sum_{i \ne j} \, e_i \,
\b{\sigma}_i \cdot \b{\sigma}_j$
and spin tensor 
${\cal T}= -B \sum_{i \ne j} \, e_i  (3 \sigma_{i\,z} \sigma_{j\,z}
- \b{\sigma}_i \cdot \b{\sigma}_j )$ have a fixed ratio. 
In the quark model this arises because
both operator structures originate from the same diagram, 
e.g., the one gluon-exchange current~\cite{Buc91}.
 But the same relation Eq.(\ref{ffrel2}) 
is obtained if we use one-pion exchange currents or a combination 
of one-gluon and one-pion exchange currents.
An evaluation of Eq.(\ref{su6break}) between $N$
and $\Delta$ spin-flavor wave functions leads straightforwardly 
to the following relations~\cite{Buc97}
\be
\label{spectroscopicquadrupolemoment}
Q_{N \to \Delta} = \frac{1}{\sqrt{2}} \, r_n^2,  
\qquad Q_{\Delta} = e_{\Delta} \, r_n^2, 
\ee 
where $e_{\Delta}$ denotes the charge of the $\Delta$ state. 

More generally, one can understand the relation between the 
neutron charge radius and the $N\to \Delta$ 
quadrupole moment in terms of a group-theoretical analysis 
without reference to specific dynamical assumptions, such as one-gluon 
exchange.  A basic assumption in a group-theoretical analysis is that
operators and states have definite transformation properties, i.e.,
they transform according to certain irreducible representations of 
the underlying 
symmetry group.   A general matrix element 
${\cal M}$ of an operator $\Omega_{R}$ 
evaluated between states belonging to the ${\bf 56}$ dimensional 
representation of SU(6) then reads 
\be
{\cal M}= \langle {\bf 56} \vert \, 
\Omega_{R} \, \vert {\bf 56} \rangle, 
\ee
where $R$ is the dimension of the irreducible representation according 
to which the 
considered operator transforms.  In the following, we denote the irreducible 
representations of the states and operators 
by their dimension. An allowed symmetry breaking 
operator $\Omega_{R}$ acting on the ground state multiplet must then transform 
according to one of the irreducible representations $R$ 
found in the product~\cite{Sak64}
\be
\label{directproduct}
\bar{{\bf 56}} \times {\bf 56} 
=  {\bf 1} + {\bf 35} + {\bf 405} + {\bf 2695}.
\ee 
Operators transforming according to other SU(6) representations
not contained in this product
will lead to vanishing matrix elements when evaluated between 
states belonging to the ${\bf 56}$.
On the right-hand side, the {\bf 1} dimensional representation is associated 
with a zero-body operator (constant), and the ${\bf 35}$, ${\bf 405}$, and 
${\bf 2695}$, are respectively connected with one-, two-, 
and three-body operators. It is important to note that one-body operators 
transforming
according to a ${\bf 35}$ dimensional representation of SU(6) do
not lift the degeneracy between $N$ and $\Delta$ masses, and do not  
generate nonvanishing baryon quadrupole moments and neutral baryon charge
radii. For these observables, first order SU(6) symmetry breaking operators 
do not suffice to obtain results in agreement with experiment.  

Higher order symmetry breaking operators can be constructed
from direct products of one-body operators.
Consequently, a general two-body spin-flavor operator transforms according 
to the irreducible representations obtained from the direct 
product ${\bf 35}\times {\bf 35}$ as follows~\cite{Sak64}
\be
\label{gsu6tbo} 
{\bf 35} \times {\bf 35} =  {\bf 1} + {\bf 35} + {\bf 35} + 
{\bf 189} + {\bf 280} + {\bar {\bf 280}} + {\bf 405}.
\ee
Two-body operators transforming according to the ${\bf 1}$  
or ${\bf 35}$ dimensional SU(6) representation can be reduced to constants 
and one-body 
operators, so that only the four higher dimensional 
representations on the right hand side of Eq.(\ref{gsu6tbo}) remain.
However, according to Eq.(\ref{directproduct}) 
only the ${\bf 405}$ appears in the direct product 
$\bar{{\bf 56}} \times {\bf 56}$. 
Therefore, within the ${\bf 56}$ an allowed two-body operator must 
necessarily transform according to the ${\bf 405}$ dimensional representation 
of SU(6). 

Next, we show that the two-body operators appearing in Eq.(\ref{su6break}) are 
components of this general SU(6) tensor. This can be seen from a decomposition 
of the tensor operator $\Omega_{\bf 405}$ into tensors with definite 
transformation 
properties with respect to the flavor and spin subgroups of SU(6) 
\bea
\label{405decomp}
{\bf 405} & = & ({\bf 1},{\bf 1}) + ({\bf 8},{\bf 1}) + ({\bf 27},{\bf 1}) 
\nonumber \\
& + &  
2 \, ({\bf 8},{\bf 3}) + ({\bf 10},{\bf 3}) +  ({\bar {\bf 10}},{\bf 3})  
({\bf 27},{\bf 3}) \nonumber \\ 
& + &   ({\bf 1},{\bf 5}) + ({\bf 8},{\bf 5}) + 
({\bf 27},{\bf 5}),  
\eea
where the first and second entry in the parentheses refers to the dimensions
of the SU(3) and SU(2) representations respectively~\cite{beg64a}. 
Because we deal with a charge operator, and the quark charge $e_i$ transforms 
as a flavor ${\bf 8}$, we confine ourselves to operators transforming as 
octets, i.e., to first order symmetry breaking in flavor 
space. Furthermore, for Coulomb multipoles, allowed operators are restricted 
to SU(2) tensors 
of even rank. Thus, the spin scalar operator ${\cal S}$ transforms as   
the $({\bf 8},{\bf 1})$ and the spin tensor operator ${\cal T}$ as 
$({\bf 8},{\bf 5})$ and both are united in a common SU(6) representation
with dimension ${\bf 405}$.  In other words, the two-body  
operators ${\cal S}$ and ${\cal T}$ in Eq.(\ref{su6break}) 
are the only allowed structures on the right hand side of Eq.(\ref{405decomp}),
and are recognized here as different components of a 
common ${\bf 405}$ dimensional tensor operator $\Omega_{\bf 405}$. 

The matrix elements of this operator evaluated between 
the ${\bf 56}$ multiplet can be factorized according to a 
generalized Wigner-Eckart theorem into 
a common reduced matrix element which is the same for the entire 
multiplet and various SU(6) Clebsch-Gordan coefficients.
The latter provide relations between the matrix elements of 
different components of the $\Omega_{\bf 405}$ and 
the ${\bf 56}$.
These group-theoretical arguments explain why the two operators ${\cal S}$ 
and ${\cal T}$  have a fixed ratio and why their matrix elements 
within the ${\bf 56}$ are related.  A derivation  of Eq.(\ref{ffrel2}) 
requires explicit SU(6) tensor representations of the operators and
states.
     
In order to further explore the range of validity 
of this relation without having to use an explicit tensor notation
for the ${\bf 405}$ and ${\bf 56}$,  
the neutron charge radius and $N\to \Delta$ transition quadrupole moment
have been evaluated in an $1/N_c$ expansion~\cite{Hes02} including 
third order SU(6) symmetry breaking due to three-body operators $\rho_{[3]}$.
The following expression has been found:
\be
\label{ffrel2statlargeN}
Q_{N \to \Delta}= \frac{1}{\sqrt{2}} \, r_n^2 \, \, 
\left ( \frac{N_c}{N_c +3} \, \sqrt{\frac{N_c+5}{N_c-1}} \right ). 
\ee
It is interesting that this more general relation 
is equivalent to Eq.(\ref{ffrel2stat}) for the physical $N_c=3$ case 
and for $N_c \to \infty$.We conclude that Eq.(\ref{ffrel2stat}) 
and its generalization to finite 
momentum
transfers in Eq.(\ref{ffrel2}) are of more general validity 
than Eq.(\ref{ffrel1}) because they also
hold in a theory~\cite{Hes02}, which includes spin-dependent three-quark terms
in the charge density, and for an arbitrary odd number of colors $N_c > 1$. 
In summary, the above arguments suggest that the relation between 
the neutron charge and $N \to \Delta$ transition form factors should be 
well satisfied in nature.

\section{Geometric shape of the nucleon}
\label{cha:3}

We have already mentioned that the relation
between the elastic and inelastic $N\to \Delta$ charge form factors 
has implications for the shape of the nucleon. The information on the
shape is contained in higher multipole moments of the nucleon charge 
distribution in particular its quadrupole moment. In this context it is 
important 
to pay attention to the coordinate frame dependence of these higher multipole 
moments. From classical electrodynamics we know that 
higher multipoles depend on the choice of origin and the orientation
of the coordinate axes.
 
\subsection{Spectroscopic and intrinsic quadrupole moments}

In ref.~\cite{Hen01} we have drawn attention to the fact that 
in order to make statements concerning the shape of the nucleon, 
one has to determine its {\it intrinsic} quadrupole 
moment. The {\it intrinsic} quadrupole moment of a nucleus
\be
\label{intrinsicqm}
Q_0=\int \! \!d^3r \, \rho({\bf r}) \,  (3 {z}^2 - {r}^2)
\ee
is defined with respect to the body-fixed frame.
If the charge density is concentrated along the $z$-direction
(symmetry axis of the particle),
the term proportional to $3{z}^2$ dominates, $Q_0$
is positive, and the particle is prolate (cigar-shaped).
If the charge density is concentrated in the equatorial plane perpendicular
to $z$, the term proportional to ${r}^2$ prevails, $Q_0$
is negative, and the particle is oblate (pancake-shaped).

\begin{figure}[htb]
\includegraphics[height=.3\textheight]{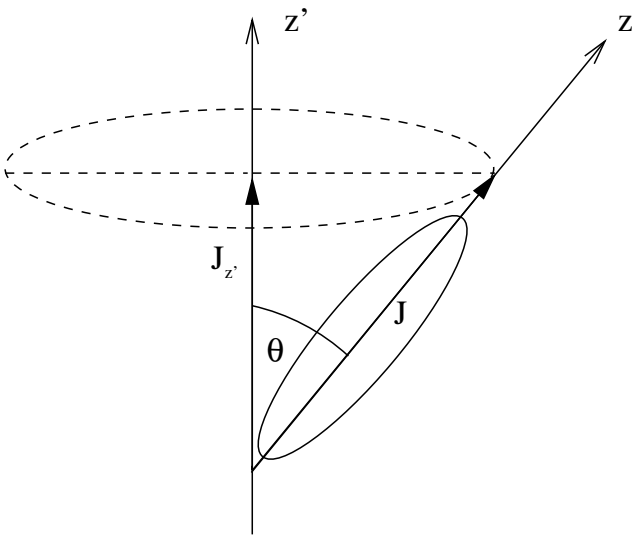}
\caption{\label{fig3}
 Precession of a deformed charge distribution  
with intrinsic symmetry axis $z$ around the laboratory frame $z'$ axis.
The transformation from the body-fixed to the 
laboratory frame gives rise to a projection factor 
$P_2(\cos(\theta))=1/2\, ( 3 \,  \cos^2(\theta) -1)$
relating spectroscopic quadrupole moment $Q$ and 
intrinsic quadrupole moment $Q_0$. Only the latter pertains to the shape of
the system.  }
\end{figure}

The intrinsic quadrupole moment $Q_0$ must be distinguished
from the {\it spectroscopic} quadrupole moment $Q$ measured in
the laboratory frame. A simple example will illustrate 
this point. Suppose one has determined the quadrupole moment $Q_0$ 
of a classical charge distribution $\rho(\bf{r})$ with symmetry axis $z$ 
and angular momentum $J$ in the body-fixed frame according 
to Eq.(\ref{intrinsicqm}).  
Then, the quadrupole moment of the same
charge distribution with respect to the laboratory frame
is given by
\bea
\label{classical}
Q= P_2(\cos \theta)\, Q_0 &  = & \frac{1}{2} \left 
(3 \, \cos^2(\theta)-1 \right ) \, Q_0 \nonumber \\ 
& = & \left ( \frac{3 {J_{z'}}^2  - J(J+1)}{2 J (J+1)} \right )
\, Q_0,
\eea
where $\theta$ is the angle between the body-fixed $z$ and the 
laboratory frame $z'$ axes, and $P_2$ is the second Legendre polynomial.  
The latter arises when tranforming the spherical harmonic 
of rank 2 in $Q_0$ from body-fixed to laboratory coordinates. 
The third equality in Eq.(\ref{classical}) is obtained 
when $\cos(\theta)$ is expressed in terms of the spin projection $J_{z'}$ 
on the laboratory frame $z'$-axis and the total spin $J$ of the system
as $\cos(\theta)=J_{z'}/\sqrt{J^2}$ (see Fig.~\ref{fig3}).  

Thus, in the laboratory one does not measure the intrinsic
quadrupole moment directly but only its projection onto the $z'$ axis.
In the quantum mechanical analogue of Eq.(\ref{classical}) 
the denominator of the projection factor 
is changed into $(2J +3)(J+1)$~\cite{Boh75}.
The projection factor shows that $J=0$ and $J=1/2$ systems
have vanishing spectroscopic quadrupole moments even though
they may be deformed and their intrinsic quadrupole moments are 
nonzero. 
%%%%%%%%%%%%%%%%%%%%%%%%%%%%%%%%%%%%%%%%%%%%%%%%%%%%%%%%%%%%%%%%%%%%
% Figure 3 Intrinsic vs. Laboratory frame
%%%%%%%%%%%%%%%%%%%%%%%%%%%%%%%%%%%%%%%%%%%%%%%%%%%%%%%%%%%%%%%%%%%%
%%%%%%%%%%%%%%%%%%%%%%%%%%%%%%%%%%%%%%%%%%%%%%%%%%%%%%%%%%%%%%%%%%%%%
For a $J=0$ nucleus this is intuitively clear because it 
does not have a preferred direction in space. All directions are equally 
probable so that in the laboratory frame a spherically symmetric charge 
distribution is observed.  A similar argument holds for the $J=1/2$ proton, 
which does not have a spectroscopic quadrupole moment and which will always 
appear spherically symmetric in the laboratory frame. 
However, the nucleon can have and does have an intrinsic quadrupole moment 
$Q_0$. It is the latter which is related to the geometric shape of the nucleon.
%%%%%%%%%%%%%%%%%%%%%%%%%%%%%%%%%%%%%%%%%%%%%%%%%%%%%%%%%%%%%

\subsection{Intrinsic quadrupole moment of the nucleon} 
  
To obtain information on the geometric shape of the nucleon 
we have used different models
to estimate its intrinsic quadrupole moment~\cite{Hen01}.
For example, in the quark model we found
\be
\label{intquad}
Q_0^p = -Q_0^{\Delta^+}= -r_n^2, 
\ee
i.e., the intrinsic quadrupole moment of the proton is equal to
the negative of the neutron charge radius and therefore
positive, whereas the $\Delta^+$ has a negative intrinsic
quadrupole moment that is equal to its spectroscopic quadrupole moment. 
This corresponds to a prolate proton and an oblate $\Delta^+$ deformation. 
A comparison of Eq.(\ref{intquad}) with Eq.(\ref{ffrel2stat}) shows that
the intrinsic quadrupole moments of the $N$ and $\Delta$ are only then
nonzero if the transition quadrupole moment $Q_{N \to \Delta}$
is nonzero.  

In the quark model, the positive intrinsic quadrupole moment of the proton 
is obtained by evaluating the two-quark charge quadrupole operator 
\be
\label{quadopqm}
Q_{[2]}  =
-B \sum_{i \ne j}^3 \, e_i 
(3 \sigma_{i\, z} \sigma_{j\,z}
- \b{\sigma}_i \cdot \b{\sigma}_j)
\ee
between quark model wave functions in which the Clebsch-Gordan coefficients 
that express the coupling of the spin 1 diquark
to the third quark have been renormalized to 1. 
The reason for doing this is the following. 
Although the two spin 1 diquarks in the proton ($uu$ and $ud$) 
have nonvanishing quadrupole moments, the coupling of the diquark 
spin to the spin 1/2 of the third quark prevents these quadrupole moments 
from being observed. After renormalization of the spin coupling coefficients, 
the contributions of the spin 1 diquarks  
in the $M=0$ and $M=1$ substates no longer cancel each other but
the $M=0$ component prevails. This results in an overall positive 
quadrupole moment, which we identify with the intrinsic quadrupole
moment of the proton $Q_0^p$. A similar calculation for the neutron shows 
that its intrinsic quadrupole moment is also positive. 
 
A simple way to picture these results is the following is to recall that 
the two-quark spin-spin operators are repulsive between quark pairs 
with spin 1. In the proton they push 
the up quarks further apart than an up-down
quark pair. This results in an elongated (prolate) charge distribution 
with the down quark in the middle.  Likewise, in the neutron 
we have an elongated (prolate) charge distribution 
with the up quark in the middle and
the negative down quarks at the periphery leading to a negative 
neutron charge radius. 
In the $\Delta^+$ with only spin 1 quark pairs there is no 
asymmetry between up-down and up-up pairs. 
This corresponds to an equilateral triangle (oblate) configuration
of its constituent quarks. However, this picture is an effective valence 
quark description 
of nucleon dynamics. In reality, the valence quarks are nearly spherically 
distributed,
and the deformation of the nucleon's charge density resides 
in the quark-antiquark pairs, which are the physical origin of the two-quark 
operators in Eq.(\ref{quadopqm}). 
In summary, the quark model with two-body exchange currents shows that 
the negative spectroscopic quadrupole moments of the $\Delta$ and the 
$N \to \Delta$ transition, and the negative neutron 
charge radius are different 
manifestations of a prolate intrinsic charge distribution of the nucleon.
	  
\begin{figure*}[htb]
\includegraphics[height=0.33\textheight]{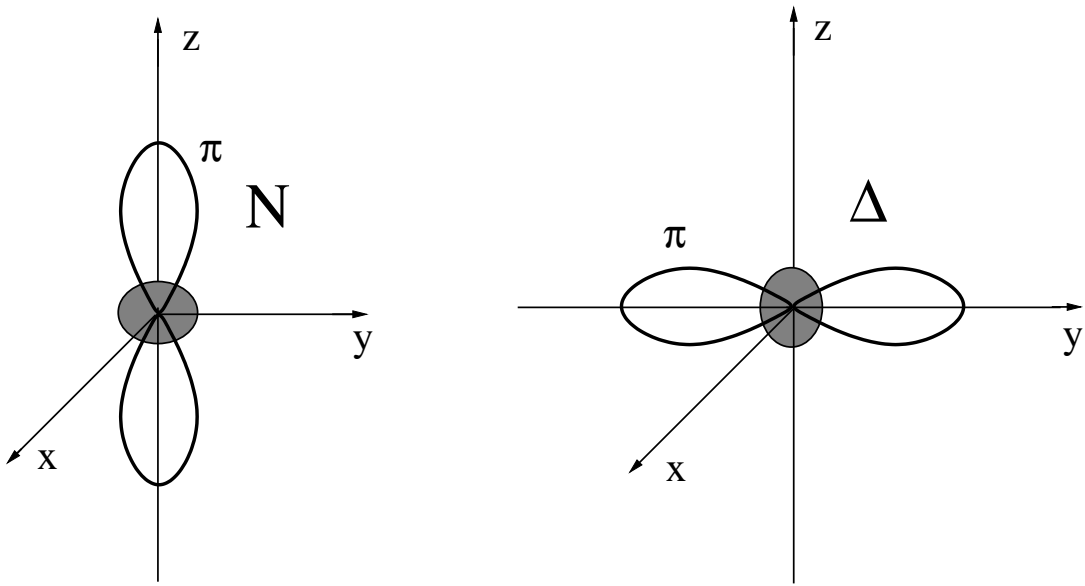}
\caption{\label{fig:pcm}
Intrinsic quadrupole deformation of the nucleon (left)
and $\Delta$ (right) in the pion cloud model. In the $N$
the $p$-wave pion cloud is concentrated along the polar (symmetry) axis,
with maximum probability of finding the pion at the poles.
This leads to a prolate deformation. In the $\Delta$, the pion cloud is
concentrated in the equatorial plane producing an oblate intrinsic
deformation. Depicted here are the 
angular ($p$-wave) parts of the pion wave functions, 
i.e. $Y^1_0$ in the case of $N$ 
and $Y^1_1$ in the case of $\Delta$ surrounding an 
almost spherical quark core (from Ref.~\cite{Hen01}). }
\end{figure*}

A quite similar result suggesting a connection between the neutron 
charge radius
$r_n^2$ and the intrinsic quadrupole moment of the proton $Q_0^p$
is also obtained in the pion cloud model (see Fig.~\ref{fig:pcm}).
In this model, the nucleon consists of a spherically symmetric
bare nucleon (quark core) surrounded by a pion with orbital angular
momentum $l=1$. 

For example, the physical proton with spin up, denoted by 
$\vert p \uparrow \rangle$, is
a coherent superposition of three different terms: (i) a 
spherical quark core contribution with spin 1/2, called a bare proton $p'$; 
(ii) a bare $p'$ surrounded by a neutral pion cloud, and 
(iii) a bare neutron $n'$ surrounded by a positively 
charged pion cloud~\cite{Hen62}. In the last two terms the spin(isospin) 
of the bare proton and of the pion cloud  are coupled to total spin and 
isospin of the physical proton.  Similarly, the physical $\Delta^+$ is
described as a superposition of a spherical quark core term with spin 3/2, 
called a bare $\Delta^{+\, '}$, a bare 
$p'$ surrounded by a $\pi^0$ cloud, and  
a bare $n'$ surrounded by a $\pi^+$ cloud.
In each term, the spin/isospin  
of the quark core and pion cloud are coupled to the total spin and isospin 
of the physical $\Delta^+$. We then write:
\begin{eqnarray} 
\label{pionwave}
\vert p \uparrow \rangle &= & \alpha 
\vert p' \uparrow \rangle 
                         + \beta 
\frac{1}{3} \Bigl (\vert p' \uparrow \pi^0 Y^1_0 \rangle 
-\sqrt{2} \vert p' \downarrow \pi^0 Y^1_1  \rangle \nonumber \\
&- &\sqrt{2} \vert n' \uparrow  \pi^+ Y^1_0  \rangle 
+ 2       \vert n' \downarrow  \pi^+ Y^1_1 \rangle \Bigr ),
\nonumber \\
%%%%%%%%%%%%%%%%%%%%%%%%%%%%%%%%%%%%%%%%%%%%%%%%%%%%%%%%%%%%%%%
%\vert n \uparrow \rangle &= & \alpha 
%\vert n' \uparrow \rangle 
%                         + \beta 
%\frac{1}{3} \Bigl (- \vert n' \uparrow \pi^0 Y^1_0 \rangle 
%+\sqrt{2} \vert n' \downarrow \pi^0 Y^1_1  \rangle 
%+\sqrt{2} \vert p' \uparrow  \pi^- Y^1_0  \rangle 
%- 2       \vert p' \downarrow  \pi^- Y^1_1 \rangle \Bigr ),
%\nonumber \\
%%%%%%%%%%%%%%%%%%%%%%%%%%%%%%%%%%%%%%%%%%%%%%%%%%%%%%%%%%%%%%%%
\vert \Delta^+ \uparrow \rangle &= & \alpha' 
\vert \Delta^{+'} \uparrow \rangle 
                         + \beta' 
\frac{1}{3} \Bigl ( 2 \vert p' \uparrow \pi^0 Y^1_0 \rangle 
+ \sqrt{2} \vert p' \downarrow \pi^0 Y^1_1  \rangle \nonumber \\ 
&+ & \sqrt{2} \vert n' \uparrow  \pi^+ Y^1_0  \rangle 
+        \vert n' \downarrow  \pi^+ Y^1_1 \rangle \Bigr ),
%%%%%%%%%%%%%%%%%%%%%%%%%%%%%%%%%%%%%%%%%%%%%%%%%%%%%%%%
%\vert \Delta^0 \uparrow \rangle &= & \alpha' 
%\vert \Delta^{0'} \uparrow \rangle 
%                         + \beta' 
%\frac{1}{3} \left ( 2 \vert n' \uparrow \pi^0 Y^1_0 \rangle 
%+ \sqrt{2} \vert n' \downarrow \pi^0 Y^1_1  \rangle 
%+ \sqrt{2} \vert p' \uparrow  \pi^- Y^1_0  \rangle 
%+        \vert p' \downarrow  \pi^- Y^1_1 \rangle \right ), 
%%%%%%%%%%%%%%%%%%%%%%%%%%%%%%%%%%%%%%%%%%%%%%%%%%%%%%%%%%%%%
\end{eqnarray}
where $\beta$ and $\beta'$ describe
the amount of pion admixture in the $N$ and $\Delta$ wave 
functions. These amplitudes satisfy the normalization conditions 
$\alpha^2 + \beta^2=\alpha^{'2} + \beta^{'2} =1$, 
so that we have only two unknows $\beta$ and
$\beta'$. 
The corresponding wave functions for the neutron and $\Delta^0$ are obtained 
by isospin rotation~\cite{Hen62}.
Here, $Y^1_0$ and $Y^1_1$ are spherical harmonics of rank 1
describing the orbital angular momentum wave functions of the pion. 
Because the pion moves predominantly in a $p$-wave,
the charge distributions of the nucleon and $\Delta$   
deviate from spherical symmetry, even if the bare nucleon and 
bare $\Delta$ wave functions are spherical. 
 
The quadrupole operator to be used in connection with these states is
\begin{equation}
\label{pionquad}
{\hat Q}={\hat Q_{\pi}} = e_{\pi} \sqrt{16 \pi \over 5} 
r_{\pi}^2 Y^2_0({\bf {\hat r}}_{\pi}),  
\end{equation}
where $e_{\pi}$ is the pion charge operator divided by the 
charge unit $e$, and $r_{\pi}$ is the distance between the center
of the quark core and the pion. Our choice of ${\hat Q}={\hat Q_{\pi}}$ implies
that the quark core is spherical and that the entire quadrupole moment 
comes from the pion p-wave orbital motion.
The $\pi^0$ terms 
do not contribute when evaluating the operator ${\hat Q}_{\pi}$ 
between the wave functions of Eq.(\ref{pionwave}).
We then obtain, e.g.,  for the 
spectroscopic $\Delta^+$ and $p \to \Delta^+$ quadrupole moments 
\be
\label{pcm1}
Q_{\Delta^+}  = -{2 \over 15} \, {\beta'}^{2}\, r_{\pi}^2, \qquad 
Q_{p \to \Delta^+}  = {4 \over 15} \, {\beta'} \beta \, r_{\pi}^2. 
\ee

%%%%%%%%%%%%%%%%%%%%%%%%%%%%%%%%%%%%%%%%%%%%%%%%%%%%%%%%%%%%%%%%%%%%
%%%%%%%%%%%%%%%%%%%%%%%%%%%%%%%%%%%%%%%%%%%%%%%%%%%%%%%%%%%%%%%%%%%%%%%%

We have to determine three parameters $\beta$, $\beta'$, 
and $r_{\pi}$. From the experimental $N \to \Delta$ quadrupole transition 
moment, $Q_{p \to \Delta^+}^{exp}\approx r_n^2 $ \cite{Bla01}, 
we can fix only one of them.
Therefore, we also calculate the 
nucleon and $\Delta$ charge radii in the pion cloud model and obtain
\bea 
\label{neutron}
r_p^2 & = & r_{p'}^2 - r_n^2,   
%-\beta^2 \, \frac{2}{3} \left ( r_{p'}^2  - r_{\pi}^2 \right ), 
\hspace{0.9 cm} 
r_n^2  =  \beta^2 \, \frac{2}{3} \left ( r_{p'}^2 - r_{\pi}^2 \right ), 
\nonumber \\
r_{\Delta^+}^2 & = & r_{p'}^2 - r_{\Delta^0}^2, 
%{\beta'}^2 \, \frac{1}{3} \left (r_{p'}^2  - r_{\pi}^2 \right ), 
\hspace{0.7 cm}
r_{\Delta^0}^2  =  {\beta'}^2 \, \frac{1}{3} 
\left (r_{p'}^2  - r_{\pi}^2 \right ).
\eea
Here, $r_{p'}^2$ is the charge radius of the bare proton. 
We have assumed that the charge radii of the bare proton and of the bare 
charged $\Delta$ states are approximately equal and that the bare neutron 
and $\Delta^0$ charge radii are zero. 
The first equation $r_{p'}^2= r_p^2 + r_n^2$ expresses the bare proton charge 
radius in terms of the experimental isoscalar nucleon charge radius. 
Subtracting the first and third equations one gets 
\be 
\label{cond}
r_p^2 - r_{\Delta^+}^2 = 
(r_{p'}^2 - r_{\pi}^2 )\left ( \frac{1}{3} {\beta'}^2 
- \frac{2}{3} {\beta}^2 \right )  = r_n^2,   
\ee
where the last equality follows if we choose $\beta' =- 2\beta$.
When the latter condition is used in Eq.(\ref{pcm1}), we get  
\be
\label{pcmsqm}
Q_{\Delta^+}= Q_{p \to \Delta^+} = r_n^2.
\ee
This is in the same ballpark as the quark model prediction of 
Eq.(\ref{spectroscopicquadrupolemoment}). 
From the experimental nucleon charge radii and Eq.(\ref{neutron}) 
one can now determine the remaining parameters $\beta$ and $r_{\pi}$ 
(see Ref.~\cite{Hen01}).  

For the spectroscopic quadrupole moment of the proton in the 
pion cloud model we obtain the following expression
\bea
\label{pcm3}
Q^p & = & {4 \over 3} \beta^2  r_{\pi}^2 \, 
\left ( \frac{1}{3} \, \langle Y^1_0 \vert P_2 \vert Y^1_0 \rangle  
+ \frac{2}{3} \,
\langle Y^1_1 \vert P_2 \vert Y^1_1 \rangle \right ) \nonumber \\
& = & 
 {4 \over 3} \beta^2  r_{\pi}^2 \, \left (
{1 \over 3} \ \left ( { 2 \over 5 } \right ) 
+  {2 \over 3} \ \left ( -{ 1 \over 5} \right ) \right )=0. 
\eea 
The factors $1/3$ and $2/3$ are the squares of the Clebsch-Gordan
coefficients that describe the angular momentum coupling of the 
bare neutron spin 1/2 with the pion orbital angular momentum $l=1$ to total
spin $J=1/2$ of the proton. They ensure that the spectroscopic
quadrupole moment of the proton is zero. The factors $2/5$ and 
$-1/5$ are the expectation values of the Legendre polynomial 
$P_2(\cos \theta)$ evaluated between the pion wave function 
$Y^1_0({\bf {\hat r} }_{\pi})$ 
(pion cloud aligned along z-axis) and
$Y^1_1({\bf {\hat r}}_{\pi})$ (pion cloud aligned along an axis 
in the x-y plane). 

In order to obtain an estimate for the intrinsic quadrupole moment 
we set {\it by hand} each of the coupling coefficients in front of 
$<Y^1_0| P_2| Y^1_0> $ and $<Y^1_1| P_2| Y^1_1> $  equal to 1/2,
thereby preserving the sum of coupling coefficients.
The cancellation between the two orientations of the cloud then disappears
and leads to an overall positive intrinsic quadrupole moment.
By this procedure we are undoing the geometric averaging over all 
angles, which prevents the nonsphericity of the pion cloud from being 
observed  in the laboratory. Furthermore, we note that the first term 
in Eq.(\ref{pcm3}), which comes from the $Y^1_0$ part of the pion wave 
function, dominates, indicating that the probability for finding the pion in
the nucleon is not spherically symmetric but larger at the poles. 
This term is just the negative of the spectroscopic $\Delta^+$ 
quadrupole moment divided by 2. We take it, properly renormalized, 
as a measure of the intrinsic quadrupole moment of the nucleon in the pion 
cloud model. One then finds for the intrinsic quadrupole moment of the proton 
and the 
$\Delta^+$  
\begin{equation}
\label{pcm4}
Q^p_0 =  {8 \over 15} \beta^2 r_{\pi}^2 = -r_{n}^2,  \qquad 
Q^{\Delta^+}_0 = r_n^2. 
\end{equation} 

Thus, the intrinsic quadrupole moment of the $p$  is positive 
and that of the $\Delta^+$ negative. They are
identical in magnitude but opposite in sign.
The positive sign of the intrinsic proton
quadrupole moment~\cite{footnote5} has a simple geometrical interpretation 
in this model. It arises because the pion is preferably
emitted along the spin (z-axis) of the nucleon (see Fig. \ref{fig:pcm}). 
Thus, the proton assumes a prolate shape.
Here, we have neglected the deformation of the bare nucleon (quark core) 
due to the pressure of the surrounding pion cloud. We emphasize that 
in this model the deformation comes only from the pion cloud and not 
from the valence quark core.
Previous investigations in a quark model with pion exchange~\cite{Ven81} 
concluded that the nucleon assumes an oblate shape under the pressure of the
surrounding pion cloud, which is strongest along the polar axis.
However, in  these studies the deformed shape of the pion cloud 
itself was ignored. Inclusion of the latter  
leads to a prolate deformation that exceeds the small
oblate quark bag deformation by a large factor.

\section{Elastic form factors and nucleon shape }
\label{cha:4} 

In this chapter we study the consequences of the relation between 
$N \to \Delta$ and elastic $N$ charge form factors for the shape of the 
nucleon in more detail. To begin with, we generalize the concept of an 
intrinsic nucleon quadrupole moment to an intrinsic quadrupole form factor 
of the nucleon. This enables us to draw further 
conclusions concerning the shape of the nucleon.

\subsection{Intrinsic quadrupole form factor of the nucleon}

The concept of an intrinsic quadrupole moment of the nucleon can 
be generalized to an intrinsic quadrupole charge distribution and a 
corresponding quadrupole form factor~\cite{buc05}. 
To show how this is done, we decompose the proton charge form factor 
in two terms,
a term resulting from a spherically symmetric charge distribution,
and a second term due to the intrinsic quadrupole deformation of the
physical charge density 
\bea
\label{voldefdecomp}
G_C^p(Q^2) & = & G_{sym}^p(Q^2) -\frac{1}{6} \, Q^2 \, G_{def}(Q^2),  
\nonumber \\
G_C^n(Q^2) & = & G_{sym}^n(Q^2) + \frac{1}{6} \, Q^2 \, G_{def}(Q^2).
\eea
The factor $Q^2$ in front of $G_{def}$ arises for dimensional reasons 
and guarantees that the normalization of the charge form factors  
is preserved. In coordinate space this corresponds to the usual multipole
decomposition of the charge density 
\be
\rho({\bf r})=\rho_0(r) Y^0_0({\bf r})
+ \rho_2(r)Y^2_0({\bf r})+ \ldots,
\ee 
where the $\rho_0$ part gives rise to $G_{sym}(Q^2)$ and the $\rho_2$ part 
is connected with $G_{def}(Q^2)$.

%%%%%%%%%%%%%%%%%%%%%%%%%%%%%%%%%%%%%%%%%%%%%%%%%%%%%%%%%%%%%%%%%%%%%%%%%%
%%%%%%%%%
\begin{center}
\begin{figure*}[t]
\includegraphics[height=0.4\textheight]{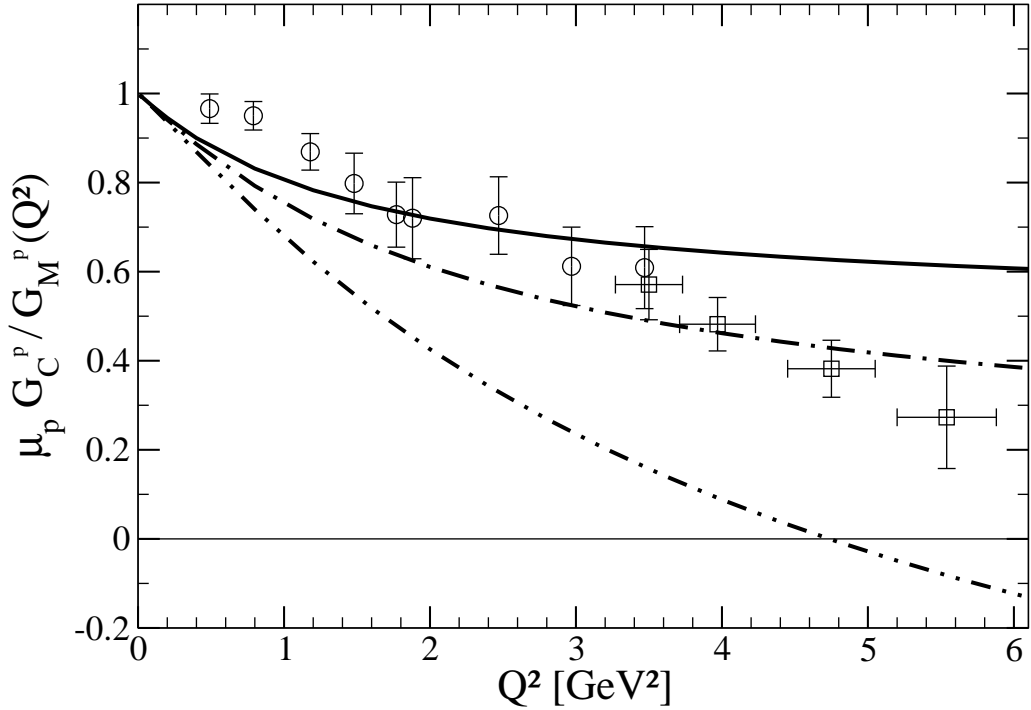}
\vspace{0.5 cm}
\caption{\label{fig:gcgm1}
Proton electromagnetic form factor ratio 
$R= \mu_p \, G_C^p(Q^2)/G_M^p(Q^2)$ as a function of
momentum transfer $Q^2$ based on the decomposition of $G_C^p$ in 
Eq.(\ref{isoscalar}). The three curves correspond to different Galster
parameters $d$ for the neutron charge form factor as in 
Fig.~\ref{fig:fig2}.The data are from
Ref.~\cite{jon00}.}
\end{figure*}
\end{center}
%%%%%%%%%%%%%%%%%%%%%%%%%%%%%%%%%%%%%%%%%%%%%%%%%%%%%%%%%%%%%%%%%%%%%%%%%%%%%
%%%%%%%%%%%

Concerning the intrinsic charge quadrupole form factor 
$G_{def}(Q^2)$, we employ the relation between the $N \to \Delta$ 
quadrupole and the elastic neutron charge form factors
in Eq.(\ref{ffrel2}), and the relation between the nucleon's intrinsic
quadrupole moment and the neutron charge radius
in Eq.(\ref{intquad}) 
\be
\label{intrinsicC2ff}
G_{def}(Q^2)= -\sqrt{2} \, G_{C2}^{N \to \Delta}(Q^2)
= \frac{6}{Q^2} \, G_{C}^n(Q^2)  
\ee
with $G_{def}(0)=-r_n^2=Q_0^p$ as discussed in chapter~\ref{cha:3}.  
This shows that Eq.(\ref{intrinsicC2ff}) is the proper generalization of the 
intrinsic quadrupole moment to finite momentum transfers.
Furthermore, we see that the deviation of the neutron charge form factor 
from zero and the deviation of the nucleon's charge distribution 
from spherical symmetry are closely related phenomena. Inserting this 
expression into Eq.(\ref{voldefdecomp}) we get
\bea
\label{isoscalar}
G_{C}^p(Q^2) & = &  G_{sym}^p(Q^2) - G_{C}^n(Q^2) = G_C^{IS}(Q^2)-G_C^n(Q^2), 
\nonumber \\
G_C^n(Q^2) & = & \frac{1}{6} \, Q^2 \, G_{def}(Q^2). 
\eea
In the case of the neutron, we find that the spherical part is zero, 
and that the neutron charge form factor is given by the nucleon's 
intrinsic quadrupole form factor. For the proton, 
the proposed decomposition implies that the spherically symmetric 
and the nonspherical intrinsic quadrupole parts are connected 
respectively with the isoscalar and the neutron charge form factors. 
Thus, the relation between the $N \to \Delta$ and neutron charge form 
factors discussed in chapter~\ref{cha:2} is seen here to have an important 
implication for the nucleon itself, which can be summarized as: 
{\it The neutron charge form factor is an 
observable manifestation and quantitative measure of 
the intrinsic quadrupole form factor of the nucleon. 
The latter also manifests itself in the proton charge form factor.} 

The decomposition suggested in Eq.(\ref{isoscalar}) 
and its interpretation can also be obtained in the pion cloud model. 
Generalizing Eq.(\ref{neutron}) to finite momentum transfers we obtain 
a decomposition of the nucleon 
charge form factors into a bare nucleon and a pion cloud part 
\bea
\label{ffpioncloud} 
G_C^p(Q^2) & = & G_C^{p'}(Q^2) + G_C^{\pi}(Q^2),  \nonumber \\
G_C^n(Q^2) & = &  - G_C^{\pi}(Q^2),  
\eea
where the bare neutron contribution to the neutron charge form factor 
is zero. Comparison of these expressions with Eq.(\ref{isoscalar}) 
shows that the bare proton contribution can be identified with 
the spherically symmetric part while the pion contribution 
is associated with the quadrupole deformation part as follows
\bea
G_C^{p'}(Q^2)  & = &  G_C^{IS}(Q^2) = G_{sym}^p(Q^2),  \nonumber \\ 
G_C^{\pi}(Q^2) & = & -G_C^n(Q^2) = -\frac{1}{6} \, Q^2 \, G_{def}(Q^2). 
\eea
In summary, both the quark model and the pion cloud model 
suggest a decomposition of the nucleon charge form factors
into a spherically symmetric quark core or bare nucleon part,   
and an intrinsic quadrupole part arising from  
quark-antiquark pairs or pion degrees of freedom.  
Furthermore, both models show that the neutron charge form factor 
is an observable manifestation of the intrinsic quadrupole charge form 
factor of the nucleon.

\subsection{Charge radii and nucleon shape}

According to the proposed decomposition of the nucleon charge form factors
in Eq.(\ref{voldefdecomp}), the nucleon charge radii can be written as a sum 
of two terms, a spherically symmetric and a nonspherical intrinsic quadrupole
contribution. The latter deformation contribution  
to the proton charge radius which makes the charge radius
bigger is given by the negative neutron charge radius
\bea
\label{deformcont}
r^2_p & = &  r^2_{p, \, sym}  +   r^2_{def}=r^2_p + r^2_n  - r^2_n,  
\nonumber \\
r^2_n  & = & r^2_{n, \, sym}  -   r^2_{def}= - r^2_{def}. 
\eea
If written in this way, we see that the spherical contribution to the proton
charge radius is given by the isoscalar charge radius
and the deformation contribution
by the negative of the neutron charge radius.
In the case of the neutron, the spherical part is zero and all of
the neutron charge radius is due to the intrinsic deformation
of its charge distribution. Eqs.(\ref{deformcont}) 
are the quark model counterparts of the pion cloud model 
formulae in Eq.(\ref{neutron}).

\subsection{Nucleon electromagnetic form factor ratio}

Nucleon recoil polarization measurements in elastic electron-proton 
scattering render it possible to extract the charge and magnetic form factors
of the nucleon with higher precision than with the Rosenbluth separation 
method that has mainly been used in the past. An important result of these 
experiments is that the proton charge over magnetic 
form factor ratio shows an almost linear decrease with increasing momentum
transfer~\cite{jon00,arr03}. In the following we show that this result can be 
understood with the help of the decomposition of the charge form factor
into two terms as discussed in the previous sections.

Using Eq.(\ref{isoscalar}) for $G_C^p$ we obtain for the proton form factor 
ratio the following expression
\bea
\label{elasticffratio}
R(Q^2) & = & 
\mu_p \frac{G_C^p(Q^2)}{G_M^p(Q^2)} \nonumber \\
& = & \frac{\mu_p}{G_M^p(Q^2)} \left (  
G_C^{IS}(Q^2) - G_C^n(Q^2) \right ).
\eea
If we insert experimental results for the isoscalar and neutron charge 
form factors, 
the measured ratio $R$ is reproduced. 
Thus, our splitting of $G_C^p$ in two terms does not modify the measured ratio 
$R$ in any way. The advantage of the proposed decomposition 
is that it provides a physical interpretation of the momentum dependence of
$R(Q^2)$. From the second equality in Eq.(\ref{elasticffratio}) we observe 
that the decrease of $R$ with increasing $Q^2$ comes from the    
intrinsic quadrupole form factor of the nucleon, which is closely 
related to the neutron charge form factor via Eq.(\ref{intrinsicC2ff}). 
This is qualitatively  
shown in Fig.~\ref{fig:gcgm1}, where we have used a simple dipole for the 
isoscalar charge and proton magnetic form factors, 
and a Galster parametrization for the neutron charge form factor. 
As mentioned above, an exact agreement between the present theory and
experiment could be established by inserting experimental data for $G_C^{IS}$ 
and $G_C^n$ instead of the above parametrizations. Nevertheless,
even the qualitative parametrizations employed here show that the observed 
decrease of $R$ can be explained by a nonspherical part in the proton charge 
distribution that is described by the neutron charge form factor. 
In coordinate space the latter leads to a more spread out charge distribution
compared to the magnetic dipole distribution~\cite{buc05,kel02}.

As clearly seen from Eq.(\ref{isoscalar}) the same quark-antiquark 
degrees of freedom, which give rise to the nonzeroness of $G_C^n(Q^2)$ and 
the related nonsphericity of the neutron's charge distribution 
also show up in the proton charge form factor. 
This is not only true at very low $Q^2$ (charge radii) and 
very high momentum transfers (decrease of $R$) but also at intermediate $Q^2$. 
Our results provide an independent confirmation of the observation made in 
Ref.~\cite{fri03} that the dip structure observed in the proton 
charge form factor at around $Q^2=0.4$ GeV$^2$~\cite{fri03} 
is due to exactly the same structure in the neutron charge form factor. 
Furthermore, according to the present theory a similar 
structure should be observed in the $N \to \Delta$ charge quadrupole 
form factor.

\section{Summary}

Advances in electron-nucleon scattering experiments have 
revealed more detailed information on the
nucleon charge form factors at high momentum transfers,
and have provided clear evidence for a small but nonvanishing 
charge quadrupole transition form factor in the excitation of 
the lowest-lying nucleon resonance $\Delta(1232)$.

Using a quark model incorporating spin-flavor symmetry and its breaking 
by spin-dependent two- and three-quark charge operators,
we have derived a relation between the neutron charge form factor 
and the $N \to \Delta$ charge quadrupole form factor. 
With the help of this relation we have expressed the 
$C2/M1(Q^2)$ ratio for $\Delta$ electroexcitation in terms 
of the elastic neutron 
form factor ratio $G_C^n/G_M^n(Q^2)$. Comparison with the data has shown that 
the two data sets satisfy the proposed relation within the experimental 
accuracy over a wide range of momentum transfers.

To draw a first conclusion concerning the geometric shape of the nucleon 
from these experiments we have pointed out that information on the shape 
of a spin 1/2 system can be obtained from its intrinsic quadrupole moment. 
Employing a quark model with exchange currents and a pion cloud model, 
we have shown that the intrinsic nucleon quadrupole moment is closely 
related to the measured $N \to \Delta$ quadrupole moment. In particular,
we have found that the nucleon's intrinsic quadrupole moment is positive and 
can be expressed by the modulus of the neutron charge radius. Thus, 
the nucleon ground state charge distribution posseses a prolate shape.

To obtain further information on the nucleon's charge distribution,
we have decomposed the nucleon charge form factors in a 
spherically symmetric and 
an intrinsic quadrupole part. For the latter we have employed the 
relation between the $N \to \Delta$ quadrupole and neutron charge form factors.
This decomposition shows that the neutron charge form factor is an observable 
manifestation of the intrinsic quadrupole deformation of the nucleon. 
In the case of the proton, the intrinsic quadrupole form factor also leads to 
observable consequences.  In particular, at high $Q^2$ it provides an 
explanation of 
the experimentally observed decrease of the proton charge over magnetic form 
factor ratio. 

As to the physical origin of nucleon deformation, all available information
suggests that collective quark-antiquark degrees of freedom, which in the 
quark model are effectively described by two- and three-quark charge 
operators, and in the pion cloud model by explicit pion degrees of freedom, 
are responsible for the nonzero neutron charge and $N\to \Delta$ quadrupole 
transition form factors, and thus for a nonzero intrinsic quadrupole 
form factor of the nucleon.  

Concerning the nucleon's magnetic form factors, we can decompose them 
into leading dipole and an intrinsic octupole part, which is related to the 
observable magnetic octupole form factor of the $\Delta$ resonance.  Because 
the latter is only nonvanishing if we include third order SU(6) symmetry 
breaking or three-quark currents, its effect on the magnetic form factors 
of the nucleon is expected to be much smaller than in the case of the charge 
form factors. We hope to discuss these matters in a forthcoming publication.

In summary, we have provided some arguments that the relation between 
the inelastic $N \to \Delta$ quadrupole and elastic neutron charge form 
factors is based on the underlying SU(6) spin-flavor symmetry of QCD,
which unites the $N$ and $\Delta$ into a common {\bf 56} 
dimensional ground state supermultiplet. In the exact symmetry limit,
$N$ and $\Delta$ masses are degenerate. In addition, the neutron charge 
form factor as well as the $\Delta$ and $N\to \Delta$ quadrupole form factors 
are exactly zero. The breaking of this symmetry by spin-flavor dependent 
two- and three-quark operators in the Hamiltonian and charge operator leads 
to the lifting of the mass degeneracy, a nonzero charge form factor of the 
neutron, as well as nonzero $\Delta$ and $N \to \Delta$ transition quadrupole 
form factors, in agreement with experiment. Moreover, the broken symmmetry 
predicts that the $N \to \Delta$ quadrupole transition form factor is related 
to the ground state neutron charge form factor. This seems to be quite well 
satisfied in nature. Finally, this relation has interesting
implications for the elastic charge form factors of the nucleon. 
It suggests that the decrease of the proton charge form factor at high $Q^2$
beyond the simple dipole behavior is due to an 
intrinsic quadrupole deformation of the proton's charge density.  
In the case of the neutron, it leads to the conclusion that 
the neutron charge form factor is a direct measure of the nucleon's intrinsic
quadrupole deformation.

\vspace{0.50 cm}
\noindent
{\bf Note added in  December 2007 \hfill}

Recently,  an analysis of the world pion electroproduction cross section 
data~\cite{dre07} has led to an empirical $C2/M1$ ratio that 
is in agreement with our prediction in Eq.(\ref{c2m1ratio}) 
also at high momentum transfers.

%%%%%%%%%%%%%%%%%%%%%%%%%%%%%%%%%%%%%%%%%%%%%%%%
%% BACKMATTER
%%%%%%%%%%%%%%%%%%%%%%%%%%%%%%%%%%%%%%%%%%%%%%%%

\begin{acknowledgments}
I would like to thank Aron Bernstein and Costas Papanicolas 
for their interest in this work.
\end{acknowledgments}



\bibliographystyle{aipproc}   % if natbib is available

\begin{thebibliography}{9}
\bibitem{comment1} The four momentum  
is defined as $Q^2=:-q_{\mu}\, q^{\mu}=\omega^2-{\bf q}^2$, 
with $q^{\mu}=(\omega, {\bf q})$ 
where $\omega=\epsilon_i-\epsilon_f$ is the energy transfer 
and ${\bf q}={\bf k}_i - {\bf k}_f$
the three-momentum transfer of the virtual photon.
Here, $(\epsilon_i,{\bf k}_i)$ and $(\epsilon_f,{\bf k}_f)$ 
are the electron energy and three-momentum in the initial
and final states.
\bibitem{comment2} Experimentally, the cross section is measured for
different incident energies $\epsilon_i$ and scattering angles $\Theta$ 
while keeping $Q^2$ and $\omega$ fixed.
The ratio $r=(d\sigma/d\Omega)_{el}/(\sigma_M \, f_{rec})$ 
is then only a function of $\tan^2(\Theta/2)$. When plotting 
the measured ratio $r$ against $\tan^2(\Theta/2)$ the resulting curve is 
a linear function with a slope given by ${G_M^N}^2$ and an intercept 
by $({G_C^N}^2 + \tau \, {G_M^N}^2)/(1 + \tau)$. The extraction of the charge 
form factor becomes increasingly more inaccurate at 
higher momentum transfers because of its suppression by the factor $\tau$.
\bibitem{comment7} The different results for the nucleon form factors obtained 
from both types of experiment can to some extent be 
understood in terms of 
two-photon exchange contributions, which affect the extraction
of the form factors from the unpolarized cross section whereas the
recoil polarization measurements are less sensitive to these 
corrections~\cite{Gui03}.
\bibitem{footnote5} After dividing by the negative
sign of the $\pi^-$ cloud, the neutron's
intrinsic quadrupole moment is also positive, i.e., $Q_0^n=Q_0^p$.
\end{thebibliography}

%%
%% End of file `template-8s.tex'.
\end{document}